\documentclass[a4paper, amsfonts, amssymb, amsmath, reprint, showkeys, nofootinbib, twoside]{revtex4}
\usepackage[english]{babel}
\usepackage[utf8]{inputenc}
\usepackage[colorinlistoftodos, color=green!40, prependcaption]{todonotes}
\usepackage{amsthm}
\usepackage{mathtools}
\usepackage{physics}
\usepackage{xcolor}
\usepackage{graphicx}
\usepackage[left=23mm,right=13mm,top=35mm,columnsep=15pt]{geometry} 
\usepackage{adjustbox}
\usepackage{placeins}
\usepackage[T1]{fontenc}
\usepackage{lipsum}
\usepackage{csquotes}
\usepackage[pdftex, pdftitle={Article}, pdfauthor={Author}]{hyperref} 
\usepackage{multirow} %
\usepackage{booktabs} %
\usepackage{float} 
\usepackage[capitalise]{cleveref}
\usepackage{hyperref} 
\hypersetup{
    colorlinks=true,
    linkcolor=black,
    filecolor=magenta,      
    urlcolor=cyan,
    pdftitle={Overleaf Example},
    pdfpagemode=FullScreen,
}
\begin{document}
\title{From yield stress to elastic instabilities: 
	\\
Tuning the extensional behavior of elastoviscoplastic fluids}

\author{Mohamed S. Abdelgawad$^\text{a}$}
\author{Simon J. Haward$^\text{b}$}
\author{Amy Q. Shen$^\text{b}$}
\author{Marco E. Rosti$^\text{a}$}\email[Correspondence email address: ]{marco.rosti@oist.jp}
\affiliation{$^\text{a}$Complex Fluids and Flows Unit, Okinawa Institute of Science and Technology Graduate University \\ $^\text{b}$Micro/Bio/Nanofluidics Unit, Okinawa Institute of Science and Technology Graduate University \\
1919-1 Tancha, Onna-son, Okinawa 904-0495, Japan}

\date{June 5, 2024} 

\begin{abstract}
In this study, we delve into the intricacies of elastoviscoplastic (EVP) fluids, particularly focusing on how polymer additives influence their extensional behavior. Our findings reveal that polymer additives significantly alter the extensional properties of the EVP fluids, such as relaxation time and extensional stresses, while having negligible impact on the shear rheology. Interestingly, the modified fluids exhibit a transition from yield stress-like behavior to viscoelastic-like behavior under high extensional rates, ultimately leading to destabilization under extreme deformation. This research enhances the fundamental understanding of EVP fluids and highlights potential advancements in applications, especially in precision-demanding fields like 3D printing.
\end{abstract}


\maketitle

\section{Introduction} \label{sec:outline}
Elucidating the complex nature of elastoviscoplastic (EVP) fluids, which embody the intriguing interplay of elasticity, viscosity, and plasticity, remains a cornerstone of non-Newtonian fluid research. These materials transition from a solid-like state below a critical yield stress to a liquid-like state above it, a behavior pivotal to the functionality of diverse substances ranging from construction materials to personal care products \cite{Balmforth2014, Coussot2014}. This transitional behavior arises from the microstructure interactions like jamming and attractive forces of colloids and gels \cite{Negro2023,Oppong2011}, extending even to larger scales in geological and extraterrestrial contexts \cite{Kostynick2022,Kurokawa2022,Dombard2006}. 
The inherent elasticity of EVP fluids extends beyond their solid state, shaping their flow characteristics, enabling energy storage and recovery post-deformation \cite{Dinkgreve2017}. This characteristic plays a crucial role in diverse phenomena, such as the loss of fore-aft symmetry and the formation of negative wakes behind sedimenting particles \cite{Holenberg2012,Fraggedakis2016}, and the inverted teardrop shape of bubbles rising through EVP fluids \cite{daneshi_frigaard_2023,Moschopoulos2021}. In practical applications, such as 3D printing, the extensional behavior of EVP fluids can enhance process efficiency \cite{Rutz2015, Rauzan2018, Zhen2020}, yet it is often accompanied by challenges like the formation of secondary droplets and elongated capillary tails that compromise precision \cite{SanTo2023}. Traditional model EVP gels used in laboratory settings, such as Carbopol and Laponite, possess limited extensibility, which contrasts with the demands of application-specific materials like printing resins.

Although the importance of extensional rheology is well-recognized in applications such as extrusion, fiber spinning, and film blowing of polymeric materials \cite{McKinley2002}, characterization of the extensional behavior of EVP fluids remains largely unexplored. EVP materials capable of substantial extensional strains open a new avenue of study, particularly given their relevance in phenomena such as pinch-off dynamics and drop formation \cite{Nelson2020}. Recent studies have focused on enhancing the extensibility of yield-stress fluids by incorporating polymeric additives into soft glassy materials, including Carbopol and oil-in-water emulsions, leading to the design of yield-stress fluids with notable extensibility \cite{Nelson2018, Nelson2019, Sen2024}. Despite these advancements, the systematic decoupling of shear and extensional rheology in such systems has not been fully realized yet, limiting their broader application potential. In addition, in works to date, the assessment of the extensional response has been based on heuristic measures such as the strain to breakup, which is not a material property but depends on imposed experimental conditions.

Here, we introduce a new EVP fluid formulation that achieves a near-complete decoupling of shear and extensional rheology through a careful selection of polymeric additives. Our model extensible EVP fluid, characterized by its thermally responsive gelation, is ideally suited for detailed flow experiments. We explore the rheological behavior of these modified fluids through comprehensive shear and extensional rheometry, revealing the effects of polymer additives on flow dynamics and normal stress behaviors. Our research contributes to the fundamental understanding and potential application enhancements of EVP fluids.  
\section{Results and Discussion} \label{sec:develop}
\subsection{Material concept}\label{subsec1}
The  model EVP fluid used in this study is a 21~wt.$\%$ aqueous solution of Pluronic F127 (PF127), which is a symmetric linear ABA triblock copolymer consisting of poly(ethylene oxide) (PEO) and poly(propylene oxide) (PPO) blocks, with chemical formula [H-(PEO)$_x$(PPO)$_y$(PEO)$_x$-OH], where $x$ and $y$ are approximately 100 and 65, respectively \cite{Prudhomme1996}. At relatively low temperatures and concentrations, PEO and PPO are both soluble in water, resulting in a solution of non-interacting individual chains of PEO-PPO-PEO molecules or unimers. As the concentration increases above the critical point of approximately 12.5~wt.$\%$, Pluronic aqueous solutions gel at a transition temperature $T_g$ near room temperature \cite{Jalaal2017}. With further increases in temperature or concentration, the hydrophobicity of the PPO block increases, leading to decreased solubility, resulting in the formation of micelles with PPO cores enclosed by a PEO shell \cite{Prudhomme1996}; see \hyperref[sec:appendix]{\textit{SI Appendix}}, \cref{fig:gelation}. The micelles consist of clusters of approximately 50 PEO-PPO-PEO chains with a core radius of approximately $4.4$~nm \cite{Prudhomme1996}. This micellar structure underlies the gelation and yield stress characteristics of PF127 solutions. The gelation is reversible and depends on the temperature. Because of their unique gelation mechanism and nontoxicity, Pluronic solutions have been extensively examined for various biological and pharmaceutical applications, including drug delivery, DNA sequencing, and cell separation \cite{Kabanov2002,Sriadibhatla2006, Cho2016, Pitto-Barry2014,Akash2015}. Additionally, they are applied in active and passive flow control in microfluidic devices due to their thermal responsiveness \cite{Bazargan2010}. Pluronic solutions have garnered attention in 3D printing applications, playing a pivotal role in the control of droplet spreading on surfaces to enhance printing accuracy \cite{Jalaal2014,Jalaal2018}. They serve as a primary component in the ink formulations of direct ink writing (DIW) systems, utilized in crafting devices for energy storage \cite{Rocha2017} and biological applications \cite{Kang2016,Kolesky2014}, due to their tunable properties and compatibility with living cells. To modify the extensional behavior of PF127 $21$~wt.$\%$, we add polymer additives of hydrolyzed polyacrylamide (HPAA). Our aim is to create an EVP material that combines the micellar structure originating from PF127 gelation, which confers yield stress, and the cross-linked structure originating from the HPAA polymer networks, which confers fluid extensibility. 

\subsection{Interior morphology}\label{subsec2}

\begin{figure}[tbp]
    \centering
    \includegraphics[width= 0.6\linewidth] {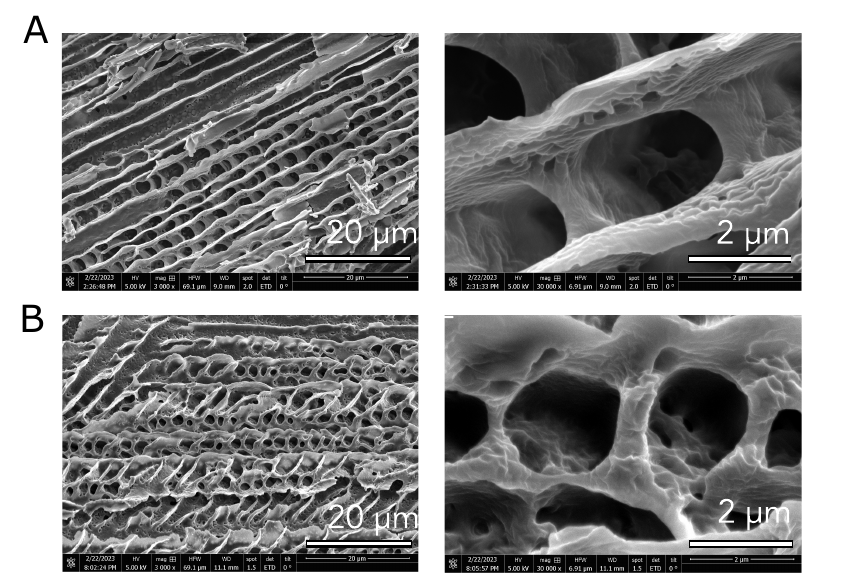}\,
     \caption{\label{fig:SEM} Scanning electron microscopy (SEM) analysis of PF127 microstructure with HPAA additive. (A) SEM image showing the intricate porous network within the freeze-dried PF127 hydrogel. (B) SEM image of PF127 with a $0.05$~wt.$\%$ HPAA additive, highlighting a reduced pore size and transition to more rounded pore morphology compared to the elongated pores in pure PF127.}
\end{figure}

A first qualitative observation of the effect of the polymer additive is achieved by looking at their scanning electron microscopy (SEM) images. The freeze-dried PF127 displays a porous structure characterized by a network of interconnected pores as shown in Fig.~\ref{fig:SEM}A. With the addition of HPAA to the PF127 solution, the SEM image shows a discernible reduction in the pore size (Fig.~\ref{fig:SEM}B). In addition, there is a notable alteration in pore morphology: the pure PF127 exhibits predominantly elongated pores with an ellipsoidal shape, whereas the addition of HPAA results in pores with a more rounded morphology. 

\begin{figure*}[tp]
    \centering
    \includegraphics[width = 1\textwidth]{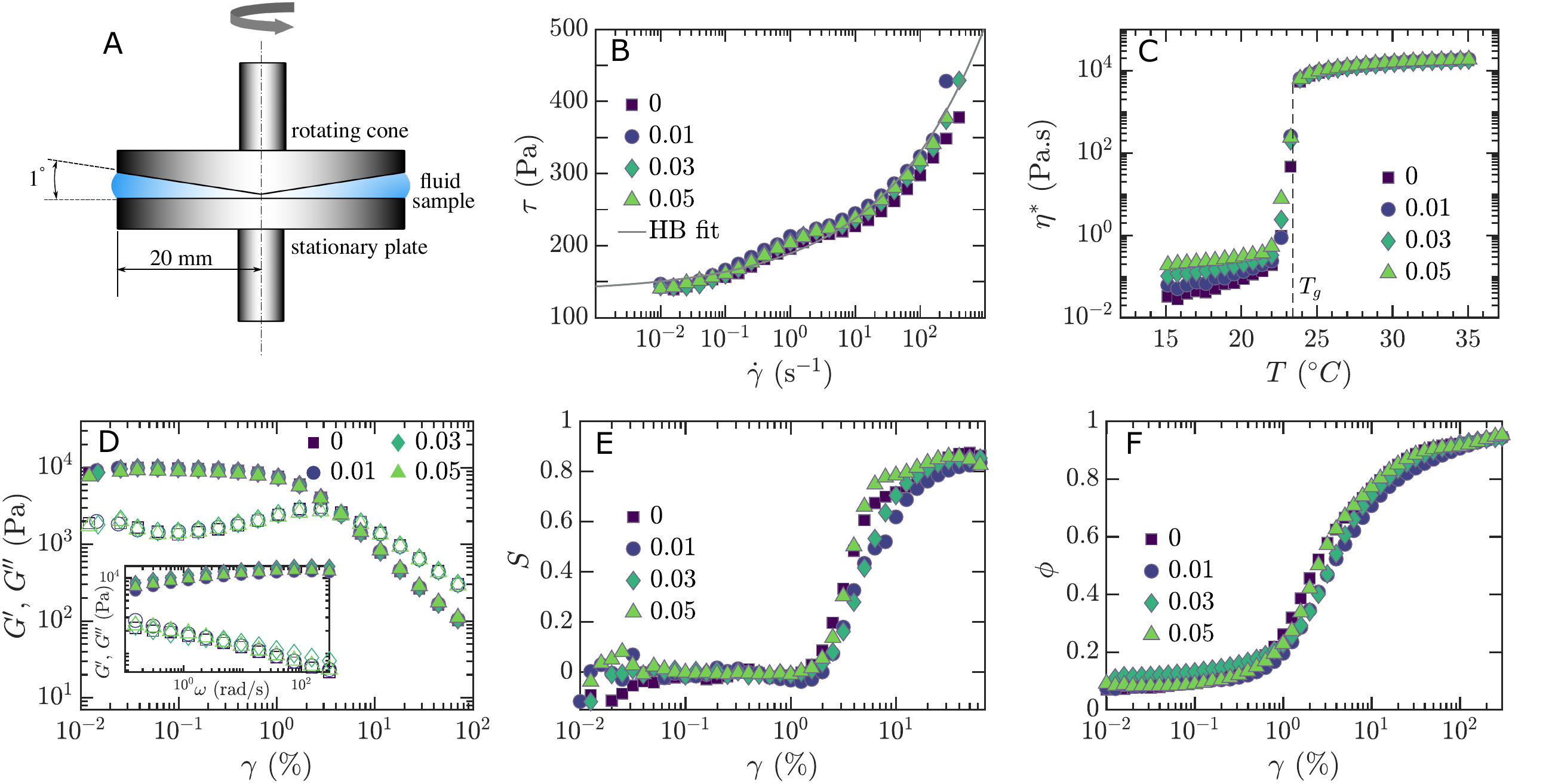}\,
     \caption{\label{fig:shear} Shear rheological properties of PF127 with HPAA additives. (A) Schematic of the shear rheometer. (B) Steady-state flow curves indicating consistent yield stress $\tau_y$ across varying concentrations of HPAA additives. The solid line represents a fit to the Herschel-Bulkley model. (C) Temperature sweeps showing the gelation temperature $T_g$, indicating no shift in gelation point due to HPAA additives. Comparative data of flow curves and temperature sweeps for PEO additives are provided in \hyperref[sec:appendix]{\textit{SI Appendix}}, \cref{fig:PEO_shear}. (D) Strain amplitude sweep at an angular frequency $\omega = 1$~rad/s (main figure) and frequency sweep at a strain amplitude $\gamma = 0.2\%$ (inset) reflecting no change in the viscoelastic moduli $G^{\prime}$ (filled symbols) and $G^{\prime \prime}$ (empty symbols) at different HPAA concentrations. (E) Strain stiffening ratio $S$, illustrating the transition from a linear elastic response at low strains ($S \approx 0$) to notable strain stiffening when $S$ increases with the strain amplitude $\gamma$ beyond the linear viscoelastic regime (LVR). (F) Energy dissipation ratio $\phi$, showing the shift from elastic behavior ($\phi \approx 0$) at smaller strain amplitudes to near-ideal plastic response ($\phi \approx 1$) at larger amplitudes. LAOS measurements in Fig. 2E-F are carried out at $\omega = 3$~rad/s. Except for Fig. 2C, the temperature is set to $26^\circ$C for all tests. Numbers in the legend of Fig. 2B-F refer to the weight percentage of HPAA additives.}
\end{figure*}

\subsection{Shear rheology}\label{subsec3}

To obtain a thorough characterization of the EVP fluid upon the introduction of polymer additives, a series of shear and extensional rheometry tests are conducted. The shear rheometry, designed to assess the impact of HPAA additives on the response of PF127 solutions to a viscometric shear flow, is  performed using a cone and plate rotational rheometer, schematically represented in Fig.~\ref{fig:shear}A.
Specifically, Fig.~\ref{fig:shear}B presents the steady-state flow curves for various HPAA concentrations. These curves reveal that the fluids exhibit a significant yield stress, a hallmark of their viscoplasticity, that interestingly does not vary with changes in HPAA concentration. To quantify the yield stress $\tau_y$, we fit the experimental data with the Herschel-Bulkley model, employing the relationship $\tau=\tau_{y}+k\dot{\gamma}^{n}$, where $\tau$ is the shear-stress obtained at the shear-rate $\dot{\gamma}$; this gives a yield stress value $\tau_y = 135\pm12$~Pa, with a flow consistency index $k = 54\pm10 $~Pa~s$^n$ and a power index $n = 0.28\pm0.04$, the latter indicating the shear-thinning behavior of these fluids. The apparent bump observed in the flow curve around $\dot{\gamma}= 10^0$~s$^{-1}$ likely results from a slip artifact leading to an underestimation of $\tau_y$. This partial slip can be eliminated or moderated by using roughened surfaces. In \hyperref[sec:appendix]{\textit{SI Appendix}}, \cref{fig:flow_curve_PP}A, we report the steady-state flow curves for pure PF127 (0~wt.$\%$ HPAA additives) using sandblasted top and bottom parallel plates (PP) for a series of different gaps to establish the gap-independent value of the yield stress. Subsequently, flow sweeps for PF127 with different concentrations of HPAA additives are conducted using a PP geometry with a 1~mm gap (\hyperref[sec:appendix]{\textit{SI Appendix}}, \cref{fig:flow_curve_PP}B). These tests result in an approximate 24$\%$ increase in the value of $\tau_y$, while confirming that the yield stress remains effectively independent of HPAA concentration. 
\\
Recent studies \cite{Thompson2018,Thompson2020} have highlighted the significance of examining the complete yield stress tensor, i.e. the stress tensor at the yield point, in yield stress materials, acknowledging the necessity to consider the effect of not just the shear component but also the normal stresses on yielding; especially when elastic effects are involved. This approach challenges the von Mises yielding criterion, based on ideal viscoplastic models, which considers the shear stress component alone since zero normal stresses are assumed, but instead considers the magnitude of the deviatoric part of the yield stress tensor that is composed of both the shear stress and the normal stress difference in simple shear flow, i.e., $\tau^{d}_{y} = \sqrt{\tau^{2}_{21,y}+1/3 (N^{2}_{1,y} + N_{1,y} N_{2,y}+N^{2}_{2,y})}$, where $\tau_{21,y}$, $N_{1,y}$, and $N_{2,y}$ are the shear stress component, first normal stress, and second normal stress at the yield point respectively. The general findings of the above-mentioned studies seem to be material specific, with most of the materials examined showing a significant $\tau^{d}_{y}$ when compared to $\tau_{21,y}$. In line with this, we report the $N_1$ and $N_1 - N_2$ in \hyperref[sec:appendix]{\textit{SI Appendix}}, \cref{fig:Normal stress differences}, measured by slowly ramping the stress around the yield point \cite{Thompson2018}, revealing no significant changes in $N_1$ and $N_1 - N_2$ across different concentrations of HPAA additives in PF127, with $\tau^{d}_{y}$ being within 5$\%$ of $\tau_{21,y}$. This consistency suggests that the magnitude of the yield stress tensor remains uniform across samples in shear and supports the validity of the von Mises yielding criterion, with a minimal contribution from normal stresses on yielding as also reported in previous studies \cite{ Habibi2016, DeCagny2019}. This indicates that our EVP fluids behave similarly to viscoplastic fluids in shear flow.
\\
Temperature ramp tests are performed to assess the influence of HPAA additives on the gelation temperature $T_g$, which demarcates the point of gel formation marked by a pronounced increase in the complex viscosity in oscillatory tests. We found that the gelation temperature remains unchanged with the addition of HPAA, and is equal to $T_g = 23.8^\circ$C, as shown in Fig.~\ref{fig:shear}C. This result suggests that HPAA additives do not significantly interfere with the gelation process of PF127. To explore the viscoelastic response of the modified fluids, strain amplitude sweeps, and frequency sweeps are conducted. The resulting material functions, namely the elastic storage modulus $G'$ and the viscous loss modulus $G''$ are observed to be mostly unaffected by the polymer additives across the examined strain amplitudes and angular frequencies (Fig.~\ref{fig:shear}D and its inset). This indicates that the inclusion of HPAA does not notably impact the elastic and viscous responses of the EVP fluids under small to large oscillatory deformations. Additionally, we apply Fourier and Chebyshev decomposition to investigate the nonlinear behavior in the large amplitude oscillatory shear (LAOS) configuration, as detailed in Ref.~\cite{Ewoldt2008,Ewoldt2010} and \hyperref[sec:appendix]{\textit{SI Appendix}}. This approach has been used extensively to reveal key rheological and structural properties critical for the functionality of food and DIW systems \cite{ALVAREZRAMIREZ2019,Garcia-Tunon2023}. Our analysis yields material functions such as the strain stiffening ratio ($S$), a measure of nonlinearity that indicates the degree of strain stiffening or softening, and the energy dissipation ratio ($\phi$), which compares the LAOS response to an ideal plastic yield-stress behavior. A value of $S=0$ denotes a linear elastic response, while positive or negative values of $S$ indicate strain stiffening or softening, respectively. The ratio $\phi$ approaches $1$ for a perfectly plastic response, zero for an elastic response, and $\frac{\pi}{4}$ for a Newtonian response. Notably, all HPAA concentrations exhibit a consistent nonlinearity measure $S \approx 0$ within the linear viscoelastic regime (i.e., for strains $\gamma \lesssim 2~\%$), and a transition to strain stiffening behavior (i.e., $S>0$) for larger strains (Fig.~\ref{fig:shear}E). The energy dissipation ratio $\phi$ shows a shift from elastic to plastic behavior as strain amplitudes increase, which is also consistent across all HPAA concentrations tested (Fig.~\ref{fig:shear}F). Additional material functions derived from LAOS experiments, including elastic and viscous Chebyshev coefficients, are detailed in \hyperref[sec:appendix]{\textit{SI Appendix}}, \cref{fig:e3_v3}. It is noteworthy to mention that, while HPAA additives show no impact on the yield stress and gelation temperature, parallel studies with PEO of of molecular weight MW = $8 \times 10^6$ Da have demonstrated a more pronounced effect. These PEO additives, at concentrations ranging from $0.1$ to $0.7$~wt.$\%$, are observed to influence the shear rheology of EVP fluids significantly (\hyperref[sec:appendix]{\textit{SI Appendix}}, \cref{fig:PEO_shear}). This difference is potentially attributable to the interactions between PEO and PF127, which may lead to an increase in the effective concentration of PF127, thereby increasing the yield stress and lowering the gelation temperature. However, given that the primary aim of this study is to modulate the extensional properties of EVP fluids, the main article focuses on HPAA additives, which do not exhibit a notable impact on shear rheology. This choice is strategic, simplifying the extensional behavior analysis and avoiding the intricacies that alterations in shear rheology might introduce.

\subsection{Extensional rheology}\label{subsec4}
\begin{figure*}[bp]
    \centering
    \includegraphics[width = 1\textwidth]{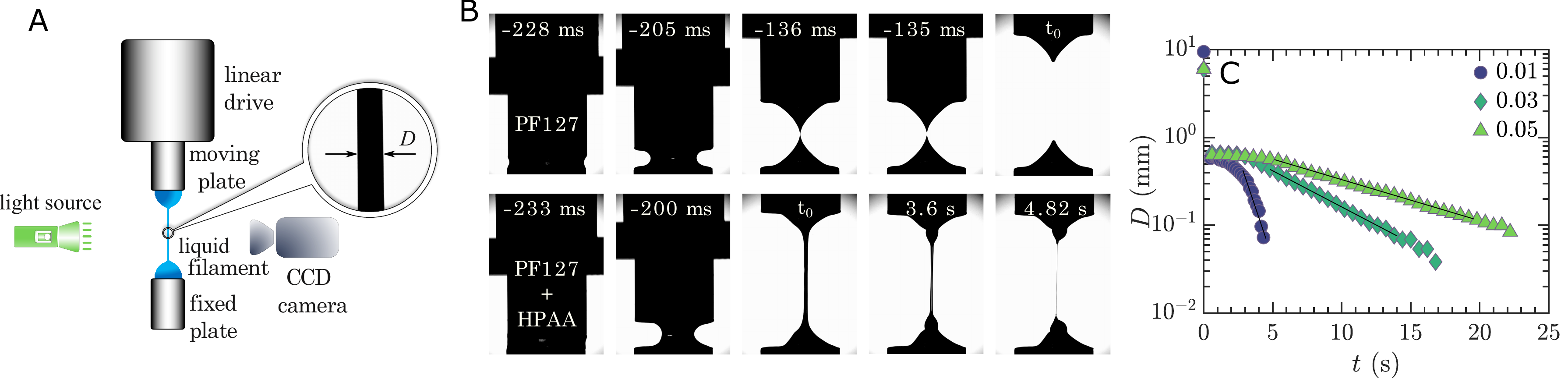}\,
     \caption{\label{fig:CaBER} Capillary break-up extensional rheometry (CaBER) of PF127 with HPAA additives. (A) Schematic of the high-speed camera and optical setup capturing filament thinning. (B) Comparison of filament behavior in PF127 without additives (top) and with $0.01$ wt.$\%$ HPAA additive (bottom). Time stamps indicate the sequence timing, with t$_0$ denoting the initial time (t $= 0$) when the upper plate reaches its final height. (C) Time evolution of the filament diameter decay, with the black lines highlighting the fitting region and legend numbers referring to the weight percentage of HPAA additives.} 
\end{figure*}

\begin{figure*}[b]
    \centering
    \includegraphics[width = 1\textwidth]{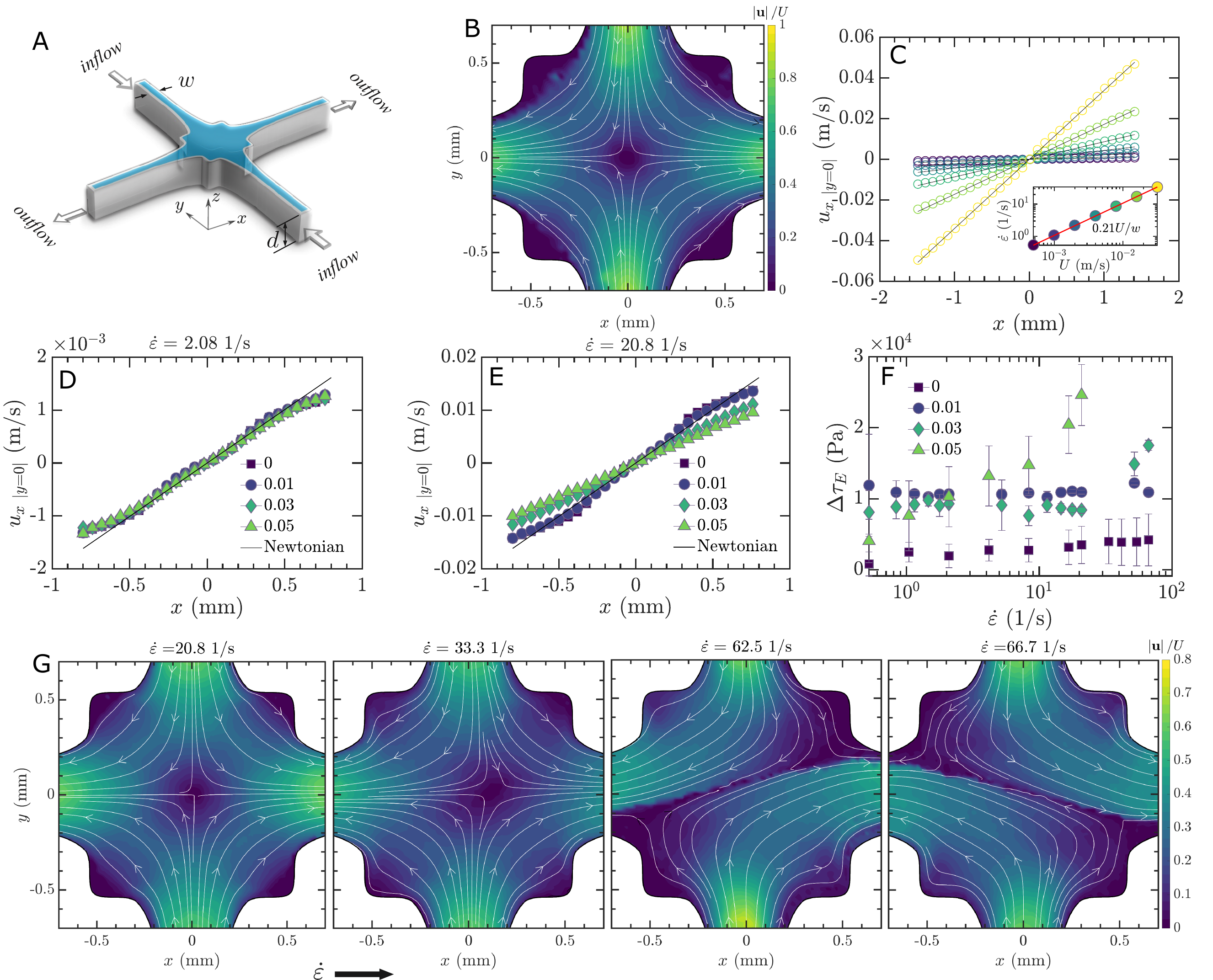}\,
     \caption{\label{fig:OSCER} Analyzing flow behaviors in EVP fluids using OSCER. (A) Schematic of the OSCER device showcasing the optimized planar cross-slot design. (B) Velocity field representation of Newtonian fluid flow, illustrating hyperbolic streamlines and a central stagnation point. (C) Velocity profiles along the $x$-axis at $y=0$, exhibiting Newtonian linearity and proportional extensional rates. (D, E) Alterations in velocity profiles for PF127 with HPAA additives at two different extensional rates $\dot{\varepsilon}$. (F) Measured extensional stress difference $\Delta\tau_E$ plotted against the imposed extensional rates $\dot{\varepsilon}$ for different HPAA additive concentrations. Numbers in the legend of Fig.~4D-F refer to the weight percentage of HPAA additives. (G) Flow fields for increasing extensional rates, showing symmetric to asymmetric transition and elastic instabilities at high rates for PF127 with a $0.05$~wt.$\%$ HPAA additive. Note that the values indicated for $\dot{\varepsilon}$ are nominal and are calculated from the velocity profiles obtained for Newtonian fluid at the same imposed flow rate
     }
\end{figure*}
To investigate the extensional properties of EVP fluids, we use a commercial capillary break-up extensional rheometer (CaBER). This device functions by placing a droplet of fluid between two plates, then rapidly separating the plates to create a liquid filament (Fig. \ref{fig:CaBER}A). The reduction in the filament diameter $D$ is tracked over time $t$ by a built-in laser micrometer. Purely viscoelastic fluids, such as aqueous solutions of PEO or HPAA, form a cylindrical filament with a uniform diameter that decreases exponentially with time as $D \sim \exp(-t/3\lambda)$, allowing for a straightforward measurements of the fluid's relaxation time $\lambda$ \cite{Entov1997,Anna2001}. On the other hand, EVP fluids do not form a perfect cylindrical filament nor exhibit uniform thinning. Consequently, capturing the thinning process accurately requires a high-speed camera coupled with an optical setup (shown in Fig.~\ref{fig:CaBER}A and detailed in the Materials and Methods section). This setup is essential for tracking the thinning at the initial point of filament narrowing, which may not occur at the mid-plane but is often observed to be above or below it \cite{Niedzwiedz2009, Niedzwiedz2010, Ng2020}. Model EVP fluids like PF127 fail to form a stable filament: they break instantly when the plates begin to separate, as depicted in the top panel of Fig.~\ref{fig:CaBER}B. The introduction of HPAA additives to PF127 enables filament formation, allowing for the recording of filament thinning over time (bottom panel in Fig.~\ref{fig:CaBER}B). By focusing on the breakup region, we present the evolution of the filament diameter in Fig.~\ref{fig:CaBER}C. Notably, the filament diameter decreases gradually over time before it starts to decay exponentially in a self-similar fashion. Relaxation times determined from three measurements at each additive concentration are $\lambda = 0.316 \pm 0.021$, $\lambda = 1.99 \pm 0.62$, and $\lambda = 2.68 \pm 0.36$ seconds for additive concentrations of $0.01$, $0.03$, and $0.05~$wt.$\%$, respectively. A summary of relaxation times measured in this study, including those for PEO additives at different concentrations, is reported in \hyperref[sec:appendix]{\textit{SI Appendix}}, \cref{table:lambda}. We also examine how polymer additives affect the extensional rheology for $T < T_g$, i.e., in the absence of the gel structure, when the fluid behaves as a viscoelastic fluid without a yield stress (\hyperref[sec:appendix]{\textit{SI Appendix}}, \cref{fig:CaBER_11C}). Notably, relaxation times for PF127 with HPAA additives show similar values for $\lambda$ regardless of temperature, suggesting that the relaxation time measured under viscoelastic conditions ($T < T_g$) can serve as a reliable indicator of the material's behavior as an extensible yield stress fluid ($T > T_g$). This finding simplifies the measurement process significantly, eliminating the need for an optical setup to capture the thinning process of EVP fluids. Other studies have employed the extensional strain-to-break as a measure of EVP fluid extensibility upon introducing polymer additives \cite{Nelson2018, Sen2024}. This method, which determines the engineering strain at which a material placed between two plates breaks under extensional stress, is typically used in solid mechanics to assess material ductility \cite{Hibbeler2003}. However, the measure of strain-to-break faces two main challenges: first, being an extrinsic property \cite{Hutchinson1978, Nelson2018}, it varies with different initial geometries and strain rates; second, it is constrained by rheometer specifications, such as maximum extensional rate and maximum final height achievable. Moreover, the direct connection between strain-to-break values and flow dynamics remains unclear. In contrast, the measure of the relaxation time ($\lambda$) is directly associated with the flow behavior, offering a more intrinsic and universally applicable measure of fluid behavior under extension.

To further elucidate the elongational properties of EVP materials and their flow behavior, we employ the optimized shape cross-slot extensional rheometer (OSCER), depicted in Fig.~\ref{fig:OSCER}A. The design of the OSCER device is based on the classic planar cross-slot geometry, where incoming flow from two opposing vertical channels converges and exits through two opposing horizontal channels. Its shape has been numerically optimized to generate an approximation to ideal planar elongational flow across a wide region around the stagnation point at the center of the OSCER geometry \cite{Haward2012}. In Newtonian fluids, streamlines near the center of the geometry mirror hyperbolic trajectories with a singular point at the geometric center (Fig.~\ref{fig:OSCER}B). Flow experiments with water as a Newtonian fluid through OSCER, using micro-particle image velocimetry ($\mu$PIV) as detailed in the Materials and Methods section, confirm the expected characteristics of the flow field over a range of flow rates used in subsequent experiments with EVP fluids. Velocity profiles along the $x$-axis, $u_x(x)$ at $y = 0$, extracted from these fields, are presented in Fig.~\ref{fig:OSCER}C; at each applied flow rate, $u_x$ is found to be proportional to $x$, indicating a constant extensional rate $\dot{\varepsilon}$ over the examined range of flow rates. The measured extensional rates linearly correlate with the average flow velocity $U$, yielding the relationship $\dot\varepsilon = 0.21U/w$ as shown in the inset of Fig.~\ref{fig:OSCER}C. This proportionality constant is in close agreement with the results from 2D numerical simulations \cite{Haward2012}, with minor variations attributable to the finite aspect ratio of the experimental OSCER device. For viscoelastic fluids, such as dilute PEO solutions, it was found that the extensional rate $\dot{\varepsilon}$ obeys Newtonian-like behavior at low flow rates, where the Weissenberg number ($Wi = \lambda \dot{\varepsilon}$) is below the critical value for the coil-stretch transition ($Wi_{c-s} = 0.5$) \cite{DeGennes1974,Larson1989}. For $Wi > 0.5$, $u_x$ remains a linear function of $x$, but a decrease in the expected strain rate can be observed. Typically, for viscoelastic fluids, the simple Newtonian-like linear dependence of $\dot\varepsilon$ with $U$ transitions to a more complex power-law dependence with increasing $Wi$ \cite{Haward2012}. Unlike the linear velocity profiles characteristic of Newtonian fluids (Fig.~\ref{fig:OSCER}C) and viscoelastic polymer solutions \cite{Haward2012}, velocity profiles measured for the PF127 EVP fluid exhibit distinctive `bumps' located at $x = \pm 0.5$~mm \cite{Varchanis2020}, as shown in Fig.~\ref{fig:OSCER}D and E for PF127 without polymer additives. These deviations have been attributed to the formation of unyielded regions within the flow, which introduce solid body rotation components, thereby disrupting the expected hyperbolic flow field. Despite these perturbations, the flow predominantly remains shear-free around the stagnation point \cite{Varchanis2020,Kordalis2021}. When polymer additives are introduced to PF127 at different concentrations, the resulting velocity profiles at low flow rates (e.g., $U = 0.002~\text{m/s}$ and $\dot{\varepsilon} = 2.1$~s$^{-1}$, as in Fig.~\ref{fig:OSCER}D) resemble those observed in pure PF127, characterized by distinct bumps. 
However, at high flow rates (e.g., $U = 0.02$~m/s and $\dot{\varepsilon} = 20.1$~s$^{-1}$, as in Fig.~\ref{fig:OSCER}E), PF127 with high concentrations of polymer additives (e.g., $0.03$ and $0.05$~wt.$\%$) begin to show viscoelastic-like behavior. This is indicated by straighter velocity profile lines without bumps, and a reduced slope (i.e., reduced $\dot\varepsilon = \partial u_x/\partial x$), which is in direct correlation with the additive concentration, reflecting the fluid relaxation time. This transition from yield stress to viscoelastic-like behavior indicates a dynamic interplay between the fluid plasticity and elasticity, contingent on the level of deformation experienced, with plasticity prevailing at low extensional rates and elasticity becoming more dominant at higher rates.

For EVP fluids with low polymer additive concentrations ($0$ and $0.01$~wt.$\%$) at all examined extensional rates ($\dot{\varepsilon}$), and for those with higher concentrations ($0.03$ and $0.05$~wt.$\%$) below certain case-dependent $\dot{\varepsilon}$ values, the velocity field remains steady, symmetric, and Newtonian-like. This is exemplified by a centrally located stagnation point from which incoming streamlines symmetrically bifurcate towards the outlet channels, as demonstrated in Fig.~\ref{fig:OSCER}G for a $0.05$~wt.$\%$ additive at $\dot{\varepsilon} = 20.8$ s$^{-1}$. With increasing $\dot{\varepsilon}$ (and consequently increasing $Wi$), the velocity field starts deviating: the stagnation point becomes laterally displaced, and the incoming streamlines bend towards this new location, indicating the onset of elastic instabilities (i.e., $\dot{\varepsilon} = 33.3$ s$^{-1}$, second panel in Fig.~\ref{fig:OSCER}G). This transition becomes more pronounced as the extensional rate further increases, particularly noticeable at $\dot{\varepsilon} = 52.1$ s$^{-1}$, where a time-dependent behavior emerges with the stagnation point constantly changing its lateral position over time, as seen in \hyperref[sec:appendix]{\textit{SI Appendix}}, Movie S1. Subsequent increases in $\dot{\varepsilon}$ lead to a more pronounced displacement of the stagnation point, resulting in a fully asymmetric steady flow field with a preferential direction, as depicted in Fig.~\ref{fig:OSCER}G for $\dot{\varepsilon} = 62.5$ and $66.7$ s$^{-1}$. This lateral displacement or preferential flow direction is arbitrary, and the flow may exhibit an opposite direction upon repetition of the experiment at the same extensional rate. Such elastic instabilities, commonly reported in viscoelastic flows within cross-slot geometries \cite{Haward2016}, are associated with intense extensional stresses at the stagnation point \cite{Poole2007}. Our observations suggest that these instabilities in our EVP fluid are predominantly due to the enhanced elastic effects, introduced by the addition of the polymer additives.

In addition to using OSCER to portray flow field differences when polymer additives are incorporated into EVP fluid, we also use it to perform extensional rheometry. This involves quantifying the extensional stress by measuring the pressure difference $\Delta P$ across an inlet and an outlet channel. The measurements are conducted in two configurations: first, we measure the total pressure difference $\Delta P_{\text{total}}$ under conditions of planar elongational flow with a central stagnation point, with all channels operational. Here, the flows must be steady and symmetric for these measurements. Next, we measure the shear stress-related pressure difference $\Delta P_{\text{shear}}$ which arises from shear at the walls, with only one inlet and one outlet channel functioning; $\Delta P_{\text{shear}}$ is typically lower than $\Delta P_{\text{total}}$ due to the absence of the extensional flow region and the associated increase in extensional viscosity. Subtracting $\Delta P_{\text{shear}}$ from $\Delta P_{\text{total}}$, we obtain $\Delta P_{\text{ext}}$, which serves as an approximation of the extensional stress difference $\Delta\tau_E$ in the flow field, i.e., $\Delta\tau_E$ $= \tau_{xx} - \tau_{yy} \approx \Delta P_{\text{ext}} = \Delta P_{\text{total}} - \Delta P_{\text{shear}}$ (see the Materials and Methods section and \hyperref[sec:appendix]{\textit{SI Appendix}}, \cref{fig:P_measure}, as well as Ref.~\cite{Haward2012}). The resulting data of $\Delta\tau_E$ as a function of the imposed extensional rate is shown in Fig.~\ref{fig:OSCER}F with each data point representing the average of three measurements and error bars indicating the standard deviation. Two key observations emerge from the results shown in Fig.~\ref{fig:OSCER}F: first, the addition of HPAA significantly enhances the extensional stress difference of the fluid compared to pure PF127. Notably, all examined concentrations of additives seem to reach a similar plateau value of $\Delta\tau_E$ at low $\dot{\varepsilon}$, indicating a consistent extensional yield stress $\Delta\tau_{E,y}$ ($\Delta\tau_E$ at the limit ${\dot\varepsilon \to 0}$) that is substantially higher than that of pure PF127. Acknowledging the experimental limitations in achieving low extensional rates and the associated error bars, it becomes evident that $\Delta\tau_{E,y}$ derived from the pressure measurements significantly exceeds the $2\tau_y$ (shear yield stress) predicted by the von Mises yielding criterion based on viscoplastic models for a planar elongational flow. Specifically, $\Delta\tau_{E,y}$ reaches $\approx 5 \tau_{y}$ (2.5 larger than predicated) for pure PF127 and escalates to $\approx 50 \tau_{y}$ (25 larger than predicated) for PF127 with HPAA additives. Factors of increase as large as 1.5 and 2.5 were previously reported for different yield stress materials \cite{Zhang2018}. These findings challenge the applicability of the von Mises yielding criterion for these fluids in complex flows despite its validity in shear flows. Second, at large $\dot{\varepsilon}$, samples with higher additive concentrations begin to diverge, exhibiting an upsurge in $\Delta\tau_E$ that aligns with the observed changes in their velocity profiles.
It is worth noting that recent studies have suggested that the effective region of extensional flow in the OSCER device is more limited than initially thought \cite{Haward2023}. In the case of Newtonian-like kinematics, a well-defined and relatively small correction factor of $\approx 0.8$ is proposed for the extensional stress difference. For viscoelastic flows at higher $\dot\varepsilon$ the correction factor is expected to increase significantly above unity and to be non-linearly dependent on the strain rate, but can only be estimated based on several assumptions of the polymeric response to the flow field. Due to the fairly small correction at low $\dot\varepsilon$ and the uncertainty involved at higher $\dot\varepsilon$ these adjustments have not been applied to the data presented in Fig.~\ref{fig:OSCER}F.

\section{Conclusion} \label{sec:conclusions}
In summary, we achieve a tunable extensional behavior of EVP fluids through the introduction of polymer additives. We establish that these additives are capable of significantly altering the extensional properties of these fluids, such as relaxation times and extensional stresses, while maintaining their intrinsic shear rheology. This enriches non-Newtonian fluid research by considering alternative model EVP fluids with high extensibility, mirroring those used in real-world applications.
It also highlights the need for new theoretical models that incorporate extensional behavior alongside shear properties, offering a comprehensive understanding of fluid dynamics relevant for industrial processes. In the context of DIW systems, printability is primarily assessed based on shear rheology \cite{Wei2023, Grosskopf2018}; factors like yield stress ($\tau_y$), shear thinning characteristics ($n$), and stiffness ($G’$) determine an ink's printability. Despite the recognized role of crosslinkers and polymer additives in optimizing ink formulations \cite{Saadi2022}, the selection has often relied on an iterative approach and use of simple characterization methods, mainly focusing on shear properties of the ink. Less attention has been given to the extensional properties of inks such as the relaxation time and its effect on the shape fidelity, durability, and overall efficacy of the printing process. The model extensible EVP fluid developed in this study provides a systematic platform for testing extensional properties in DIW systems, whether used as inks or as matrices in embedded 3D printing, offering a new dimension of control over material behavior for improved precision and efficiency.
Our study reveals a critical transition in the behavior of EVP fluid from yield stress-like to viscoelastic-like under increased extensional rates, which impacts their stability under significant deformation, leading to elastic instabilities and chaotic behaviors. These insights pave the way for enhanced application techniques in industries where fluid dynamics play an essential role in processes such as mixing, heat transfer, and enhanced oil recovery.
\section{Materials and methods} \label{sec4}
\subsection{Material preparation}\label{subsec4-1}
A Pluronic F127 (PF127) aqueous solution of $21$~wt.$\%$ is prepared by gently mixing the Pluronic powder (Sigma-Aldrich) of molecular weight MW = $12.6$~kDa, without prior purification, into Milli-Q water using a magnetic stirrer. This mixture is stirred continuously for $24$~hours at a temperature of approximately $4^\circ$C and subsequently allowed to equilibrate for $48$ hours within a refrigeration unit before usage or adding the polymer additive. The polymer additive used in this study is hydrolyzed polyacrylamide (HPAA) with MW $= 18 \times 10^6$~Da, at concentrations of $0$, $0.01$, $0.03$, and $0.05$~wt.$\%$. The polymer additive is incorporated into the PF127 solution and mixed thoroughly using a roller mixer in a cold environment maintained at $4^\circ$C. 

\subsection{Scanning electron microscopy}\label{sec4-2}
To characterize the structure and morphology of the samples, scanning electron microscopy (SEM) (Helios NanoLab 650, Thermo Fisher Scientific) is used. Gelled samples of $1$~mL are frozen in liquid nitrogen to create mechanically fractured surfaces. Samples are then freeze-dried in a vacuum chamber for $24$~hours to ensure that the solvent has been removed and the sample is completely dry, leaving its microporosities undamaged. The dry samples are mounted on metallic stubs and coated with gold using the ion sputter-coater (MC1000, Hitachi) before being subjected to electron beam scanning for visualization.

\subsection{Shear rheology}\label{sec4-3}
Shear rheology measurements are performed using a stress-controlled DHR-3 rotational rheometer (TA Instruments, Inc.) equipped with a $40$~mm diameter and $1^\circ$ cone and plate geometry. The sample is loaded into the rheometer shortly after being taken out from the refrigerator, and the temperature upon loading is precisely regulated using a Peltier bottom plate and held at $10^\circ$C to ensure that the sample remains in the solution state during loading \cite{Hopkins2019}. A sandblasted bottom plate is used to mitigate the slip effect. For effective evaporation, humidity, and uniform temperature distribution control, we use a cone geometry with a solvent well-filled with water together with a solvent trap covering the sample throughout the measurement process.

The sample is loaded into the rheometer in a solution state and heated to $26^\circ$C for $10$~min to reach thermal equilibrium. To remove any start-up effects, the sample is then presheared at $0.01$~s$^{-1}$ for $5$~min before directly starting the flow sweep test. The shear rate is increased logarithmically in steps from $0.01$ to $500$~s$^{-1}$ and then decreased back to $0.01~\text{s}^{-1}$ in the same manner. The waiting time for each measurement was not explicitly specified. However, by enabling the steady-state sensing algorithm, the rheometer monitors the time evolution of the dependent variable in the shear test and determines when a steady state has been reached.

Prior to conducting oscillatory shear tests, the sample is subjected to the same loading and resting procedures as described for the flow sweep. In the strain amplitude sweep test, the oscillatory shear is applied at a fixed angular frequency of $1$~rad/s, increasing the strain amplitude logarithmically from $0.01\%$ to $100\%$. At a strain amplitude of $0.2\%$, within the linear viscoelastic regime, the frequency is swept logarithmically from $300$ to $0.1$~rad/s. The gelation temperature is determined through temperature sweeps at a heating rate of $0.5^\circ$C/min under an oscillatory strain amplitude of $0.5\%$ and an angular frequency of $1$~rad/s.

Large-amplitude oscillatory shear (LAOS) tests are performed using the strain-controlled rheometer ARES G2 (TA Instruments, Inc.) with a 1~mm gap, 40~mm stainless steel sandblasted parallel plate equipped with a solvent trap. Strain amplitude sweeps are carried out in two modes: correlation mode (logarithmic sweep, 10~points per decade) and transient mode (1024~points per cycle, 5 delay cycles, and six half cycles) at strains ranging from $0.01\%$ to $300\%$ with a fixed frequency of $3$~rad/s at $26^\circ$C.

\subsection{Extensional rheometry}\label{sec4-5}
\subsubsection{CaBER}\label{sec4-5-1}
Extensional rheology experiments are carried out using a capillary breakup extensional rheometer (HAAKE CaBER-1, Thermo Scientific) equipped with a high-speed camera setup and telecentric light source, allowing the capture of the filament thinning process over time; see Fig.~\ref{fig:CaBER}A. For easy loading, the sample is first loaded in a solution state at $\approx15^\circ$C and then allowed to reach the measuring temperature of $26^\circ$C using a temperature controller before doing the experiments. The initial and final plate separation distances are set at $1.0$~mm and $19.7$~mm, respectively, corresponding to a Hencky strain of $2.98$. The plates are separated using a cushioned stretching profile within 40~ms, and the thinning process is recorded at $250$ to $1000$~frames per second. The images are analyzed with a custom Matlab code to deduce filament diameter over time and relaxation time. Our optical system is validated by comparing the relaxation time for a pure viscoelastic solution of PEO $0.5$~wt.$\%$ obtained by the built-in laser micrometer in the commercial CaBER instrument with that obtained through image analysis. An excellent agreement between the two measurements is obtained, and the results are presented in \hyperref[sec:appendix]{\textit{SI Appendix}}, \cref{fig:CaBER_valid}.  

\subsubsection{OSCER}\label{sec4-5-2}
As previously mentioned, Pluronic exists in a solution state and behaves like a Newtonian fluid at a temperature below its gelation temperature. Hence, the Pluronic solution is loaded into the OSCER device and the attached tubing and syringes in a cold storage room at $\approx 4^\circ$C. The setup is then transferred to the experimental space (set at $26^\circ$C) and kept for enough time to reach thermal equilibrium with the surroundings and, consequently, turn to a gel state before carrying out any measurements. The flow inside the device is driven at a constant flow rate by a syringe pump (neMESYS, Cetoni GmbH), driving two $25$~ml glass syringes (SGE, Trajan), with one connected via a T-junction to the two inlets and the other connected via a T-junction to the two outlets of the OSCER device, in a push-pull configuration. The OSCER device is fabricated from stainless steel and features glass windows for flow visualization \cite{Haward2012}. To minimize hydraulic compliance within the system, the syringes are connected to stainless steel Swagelok tubing using luer lock fittings, leading to the four ports (two inlets and two outlets) of the OSCER geometry.

Prior to $\mu$PIV measurements, the Pluronic solution is seeded with $2~\mu$m fluorescent polystyrene particles (PS-FluoRed-Particles; Microparticles GmbH) to a concentration of $\approx 0.02$~wt$\%$ before loading. The particles are excited using a volume illumination system (TSI Inc.) with a $527$~nm dual-pulsed Nd:YLF laser installed on an inverted microscope (Nikon ECLIPSE Ti-S) equipped with a $10$X objective lens (Nikon PlanFluor). The $\mu$PIV images are captured by a $1280 \times 800$~pixels high-speed CMOS camera (Phantom Miro M310, Vision Research Inc.) with a frame rate that depends on the imposed flow rate. The fluid is driven through the OSCER device using syringe pumps over a controlled volumetric flow rate range of $0.025 < Q < 3.2$~mL/min. The average flow velocity is calculated using the formula $U = Q/wd$, where $w$ and $d$ are the width and the depth of the OSCER and have values of $200$~$\mu$m and $2.1$~mm, respectively. The nominal extensional rate is then calculated as $\dot{\varepsilon} = 0.21 U/w$ \cite{Haward2012}. The time interval between laser pulses is adjusted according to the imposed flow rate such that the particles are allowed to travel a maximum distance of $\approx 8$~pixels between two consecutive frames of image pairs. The device is placed such that the $x-y$ plane is perpendicular to the light source. The device mid-plane is brought to focus, and $100$ image pairs are captured and processed using an ensemble average cross-correlation PIV algorithm (TSI Insight 4G) to obtain the velocity field. 

A differential pressure sensor (GE Druck) with a pressure scale of $150$~psi and a sensitivity of $0.01\%$ of its full scale has been installed with the high-pressure port located near an inlet and the low-pressure port positioned at an outlet. At each imposed flow rate, the pressure drop is measured once a steady-state pressure has been achieved. Additional information regarding the pressure drop measurements is available in the \hyperref[sec:appendix]{\textit{SI Appendix}}, \cref{fig:P_measure}.
\section*{Acknowledgements} \label{sec:acknowledgements}
The authors thank the OIST Imaging Section for providing access to the scanning electron microscopy equipment. M.S.A. thanks Dr. Cameron Hopkins for his initial support and insightful discussions.

\section*{Funding} \label{sec:funding}
The research was supported by the Okinawa Institute of Science and Technology Graduate University (OIST) with subsidy funding from the Cabinet Office, Government of Japan. S.J.H acknowledges funding from the Japan Society for the Promotion of Science (JSPS), grant 21K03884.

\appendix*
\section{Supplementary information} \label{sec:appendix}
\setcounter{equation}{0}
\setcounter{figure}{0}
\setcounter{table}{0}
\makeatletter
\renewcommand{\theequation}{S\arabic{equation}}
\renewcommand{\thefigure}{S\arabic{figure}}
\renewcommand{\thetable}{S\arabic{table}}
\subsection*{Large-amplitude oscillatory shear (LAOS) measurements}
Fourier Transform (FT) rheology allows for the quantification of nonlinearities in LAOS by converting the time-domain stress response into a frequency spectrum, identifying intensities and phases of higher harmonics. In particular, given a sinusoidal strain input, $\gamma (t) = \gamma_0 \sin(\omega t)$, the stress response is expressed as
\begin{equation}
\tau(t;\omega,\gamma_0 ) = \gamma_0 \sum_{n : odd} \left\{ G_n'(\omega,\gamma_0) \sin(n\omega t) + G_n''(\omega,\gamma_0) \cos(n\omega t) \right\}.
\end{equation}
Within the linear viscoelastic regime (LVR), the response includes only the first harmonic ($n=1$), and $G_1'$, $G_1''$ are the storage and loss moduli, respectively. Nonlinear responses are signified by the presence of higher harmonics, which increase with strain and manifest in the stress response, producing unique FT spectra with peak intensities at odd harmonics (Fig.~\ref{fig:StreesResponse_I}). These can be interpreted using higher harmonics coefficients $G_n'$ and $G_n''$, providing a detailed view of the material nonlinearities beyond the first harmonic information adopted in the correlation mode \cite{Garcia-Tunon2023}.

Although FT rheology is sensitive in detecting the nonlinear response, it lacks a physical interpretation of these higher harmonics. Ewoldt et al. \cite{Ewoldt2008,Ewoldt2010} developed a framework combining FT and Chebyshev polynomials to analyze nonlinear viscoelasticity. Utilizing Chebyshev decomposition, this approach allows for a physical interpretation of the higher-order coefficients. The elastic ($e_n$) and viscous ($v_n$) Chebyshev coefficients are calculated from the Fourier coefficients as
\begin{equation}
e_n = G_n'(-1)^{\frac{n-1}{2}} \quad n: odd,  
\end{equation}
and
\begin{equation}
v_n = \frac{G_n''}{\omega} = \eta_n' \quad n: odd.
\end{equation}
Positive and negative values for these coefficients imply different physical behaviors. A positive $e_3$ indicates intracycle strain stiffening, while a negative $e_3$ corresponds to strain softening. Similarly, positive $v_3$ denotes shear-thickening and negative $v_3$ shear-thinning. The nonlinear elastic moduli, $G_L'$ and $G_M'$ (Eq.~\ref{eq:Gl} and \ref{eq:GM}), are derived from the Fourier and Chebyshev coefficients and based on their definition from the Lissajous–Bowditch (L–B) curve (Fig. \ref{fig:S_phi_def2}A). They offer a measure of the strain stiffening ($G_L' > G_M'$) or softening ($G_L' < G_M'$) within cycles;
\begin{equation}
\label{eq:Gl}
G_L' \equiv \frac{\tau}{\gamma}|_{\gamma=\pm\gamma_0} = \sum_{n: odd} G_n'(-1)^{\frac{n-1}{2}} = e_1 + e_3 + \cdots,
\end{equation}
and
\begin{equation}
\label{eq:GM}
G_M' \equiv \left. \frac{d\tau}{d\gamma} \right|_{\gamma=0} = \sum_{n: odd} n G_n' = e_1 - 3e_3 + \cdots.
\end{equation}
The strain stiffening ratio $S$, defined as 
\begin{equation}
S \equiv \frac{G_L' - G_M'}{G_L'},
\end{equation}
is a nonlinearity index, approaching zero for linear responses, being positive for stiffening, and negative for softening materials.

The perfect plastic dissipation ratio, $\phi$, quantifies the proximity of LAOS responses to ideal plastic yield-stress behavior, comparing the energy dissipated in a cycle (the shaded area in the L-B curve shown in Fig.~\ref{fig:S_phi_def2}A) to the energy that would be dissipated in an equivalent perfect plastic response (the area enclosed by the gray rectangle in Fig.~\ref{fig:S_phi_def2}B). $\phi$ approaches 1 for a perfect plastic behaviour, 0 for an elastic one, and $\frac{\pi}{4}$ for a Newtonian behavior. It can be calculated using the first-order viscous Fourier coefficient as: 
\begin{equation}
\phi = \frac{\pi \gamma_0 G_1''}{4\tau_{max}}.
\end{equation}

Our FT analysis is validated by comparing the first harmonic moduli, $G_1'$ and $G_1''$, and the stress, $\tau$, calculated from the Fourier coefficients against those obtained by the TRIOS software in correlation mode, as illustrated in Fig.~\ref{fig:LAOS_valid} for PF127. The agreement is excellent, and for conciseness, we do not include this comparison for the other samples containing HPAA additives. Across all HPAA concentrations, no notable differences are observed in the LVR width, where nonlinearity measures $\frac{e_3}{e_1}$, $\frac{v_3}{1}$, and $S$ are approximately zero. As the strain amplitude ($\gamma_0$) increases beyond the LVR, the elastic nonlinearity measure $\frac{e_3}{e_1}$ (Fig.~\ref{fig:e3_v3}A) and the strain stiffening ratio $S$ (Fig.~2E, main article) become positive and rise sharply, signifying strain-stiffening behavior. Conversely, $\frac{v_3}{v_1}$ (Fig.~\ref{fig:e3_v3}B) initially increases, indicating shear-thickening, before decreasing at higher strain amplitudes, suggesting a transition from shear-thickening to shear-thinning as the amplitude increases. Observations of the perfect plastic dissipation ratio, $\phi$, suggest nearly purely elastic behavior at small amplitudes, where $\phi$ is close to zero, transitioning to ideal plastic behavior at large amplitudes as $\phi$ approaches unity (Fig.~2F, main article). 

\begin{figure}[H]
\centering
\includegraphics[width=0.9\textwidth]{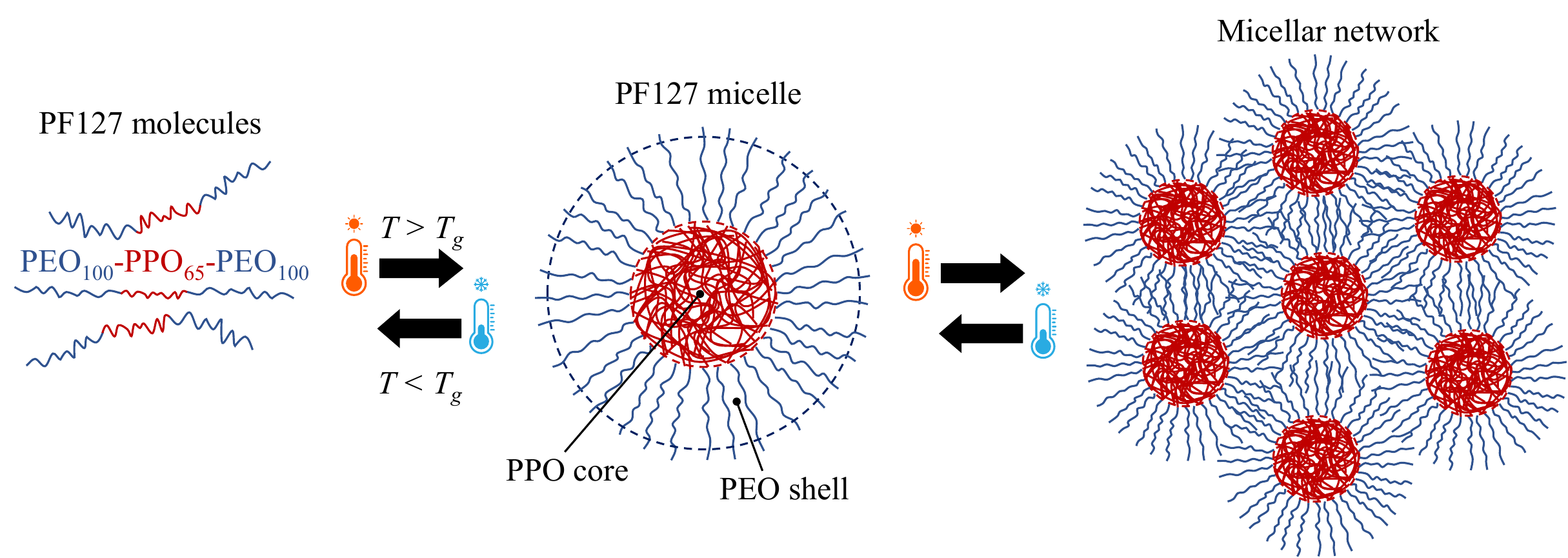}
\caption{\label{fig:gelation}Schematic representation of the thermal responsive gelation mechanism of PF127.}
\end{figure}

\begin{figure}[H]
\centering
\includegraphics[width=0.9\textwidth]{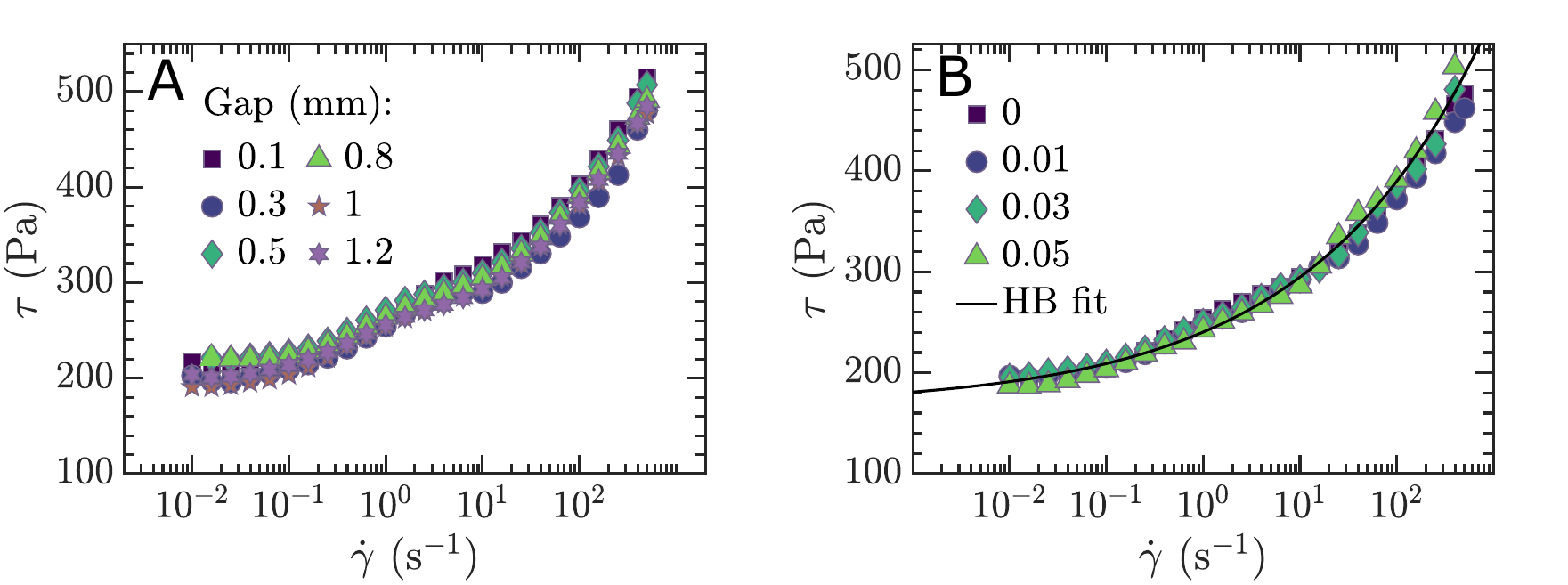}
\caption{\label{fig:flow_curve_PP} Steady-state flow curves obtained using parallel plate geometry: (A) Pure PF127 with 0~wt.$\%$ HPAA additives, tested at various gap sizes, and (B) PF127 with varying concentrations of HPAA additives (indicated in the legend), each measured with a 1 mm gap. The solid line represents a fit to the Herschel-Bulkley model that gives $\tau_y = 167.1$~Pa, $k = 73$~Pa~s$^n$ and $n = 0.24$
}
\end{figure}

\begin{figure}[H]
\centering
\includegraphics[width=0.9\textwidth]{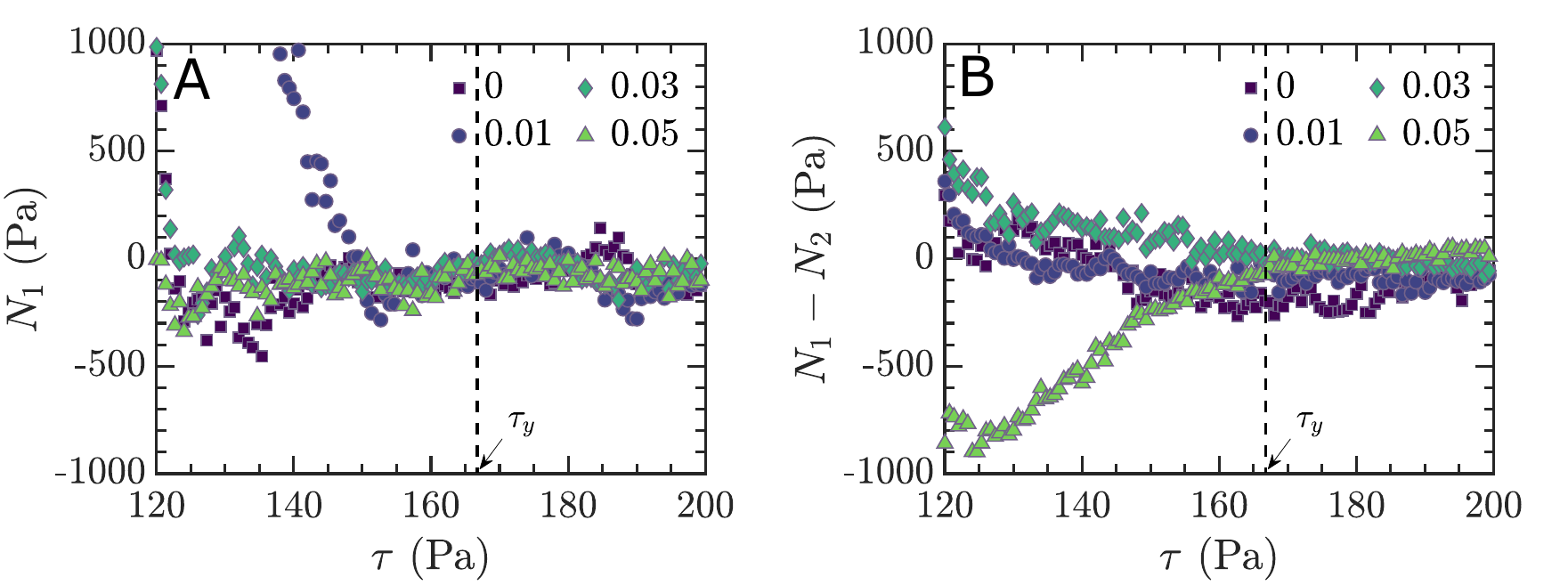}
\caption{\label{fig:Normal stress differences} Normal stress differences. (A) First normal stress difference ($N_1$) measured using a 40~mm and $1^\circ$ cone and plate geometry. 
(B) Difference between the first and second normal stress differences ($N_1 - N_2$) measured using a 40~mm parallel plate geometry with a 1~mm gap. Both are plotted as functions of shear stress near the yield point. Samples were presheared at $0.01$~s$^{-1}$ for 5 minutes, allowed to rest for 20 minutes, then subjected to a stress ramp from 120 to 200~Pa at a rate of 1~Pa/min. The vertical dashed line indicates the yield stress ($\tau_y = 167.1$~Pa), derived from parallel plate measurements with a 1~mm gap as shown in Fig. \ref{fig:flow_curve_PP}B. At the yield point, the $N_{1,y}$ and $N_{2,y}$ values obtained from these measurements are $-100$, $-63$, $-46$, $6$ and $119$, $94$, $-2$, $3$ for 0, 0.01, 0.03, and 0.05 wt.$\%$ HPAA additives respectively. The numbers in the legend denote the weight percentages of HPAA additives.
}
\end{figure}

\begin{figure}[H]
\centering
\includegraphics[width=0.9\textwidth]{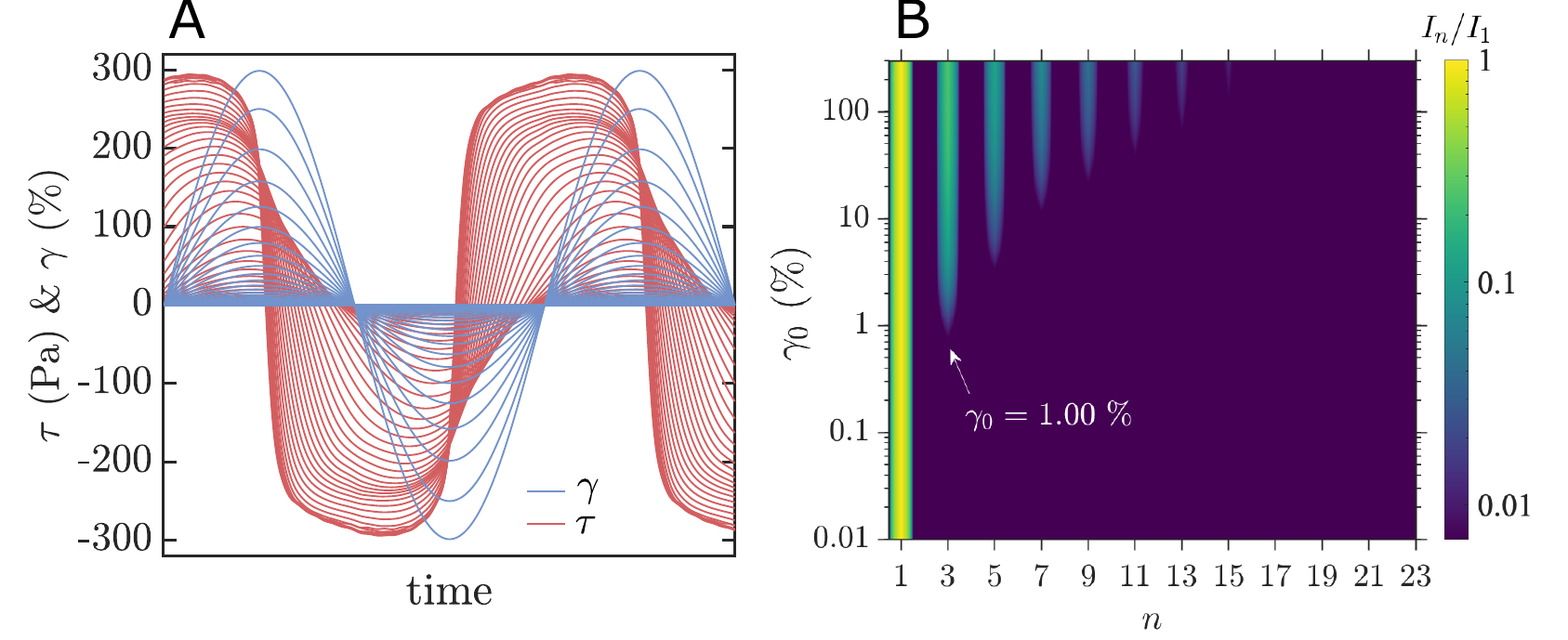}
\caption{\label{fig:StreesResponse_I}Fourier transform analysis of stress response in pure PF127. (A) Stress response (red) to sinusoidal strain input (blue) across the amplitude sweep, showing distortion at high strain amplitudes due to higher harmonic contributions. (B) Contour plot illustrating the relative intensity of odd harmonics $I_n/I_1$ from the Fourier analysis. The arrow marks the strain amplitude $\gamma_0$ at which nonlinearities become significant, indicated by $I_3/I_1$ exceeding $0.01$.}
\end{figure}

\begin{figure}[H]
\label{fig:S_phi_def1}
\centering
\includegraphics[width=0.9\textwidth]{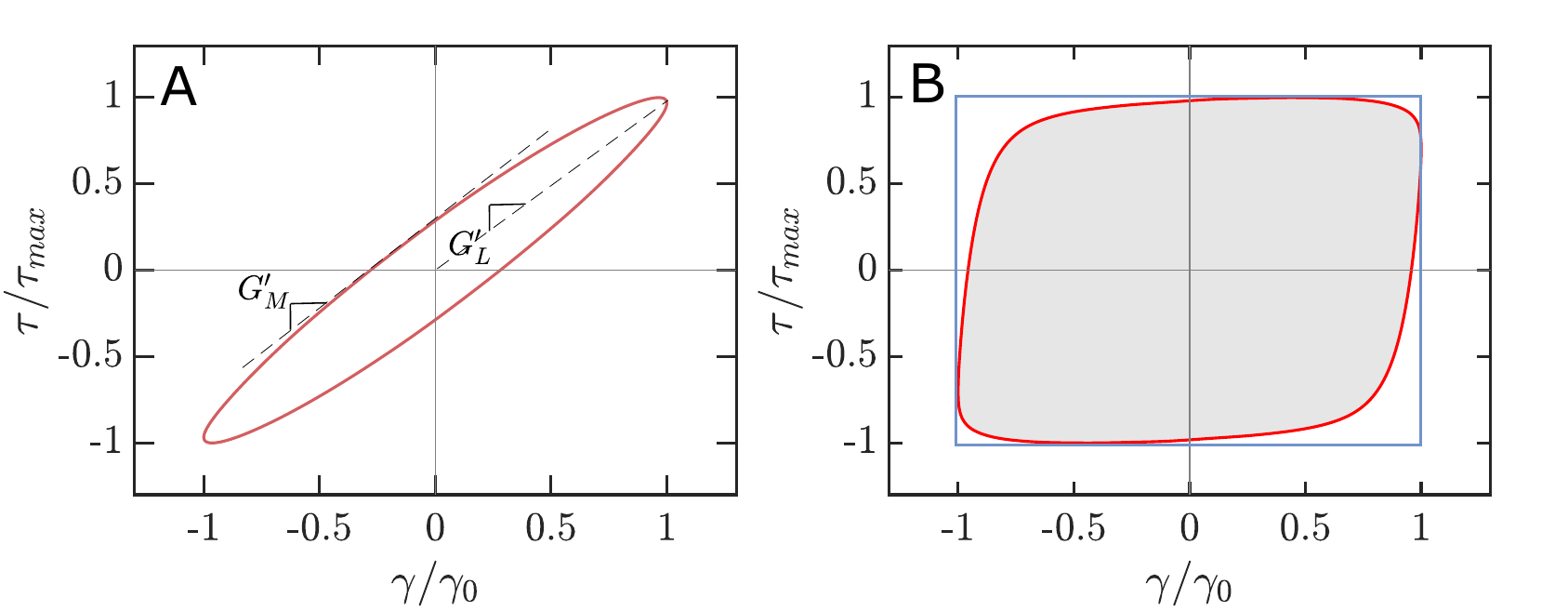}
\caption{\label{fig:S_phi_def2}Geometric interpretation of nonlinear elastic moduli via Lissajous–Bowditch curves. (A) Geometric illustration defining the nonlinear elastic moduli, $G_L'$ and $G_M'$, (Eq.~\ref{eq:Gl} and \ref{eq:GM}). (B) Shaded area illustrates the energy dissipated within a LAOS cycle, while the blue rectangle represents the energy dissipation expected from a perfect plastic material under identical strain conditions.}
\end{figure}

\begin{figure}[H]
\centering
\includegraphics[width=0.9\textwidth]{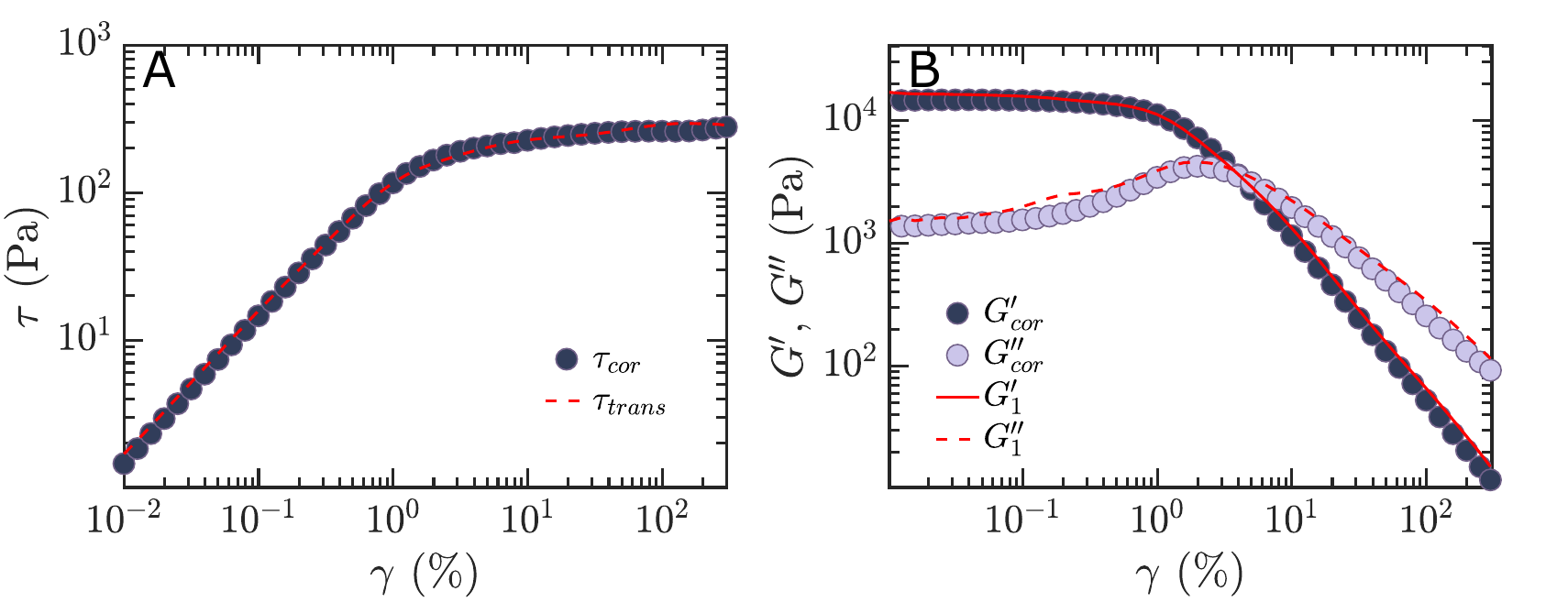}
\caption{\label{fig:LAOS_valid}Validation of Fourier transform rheology analysis for pure PF127. (A) Stress amplitude versus strain amplitude from correlation (cor, symbol) and transient (trans, dashed line) data. (B) Comparison of the first harmonic moduli, $G_1'$ (solid line) and $G_1''$ (dashed line), calculated from Fourier coefficients with the corresponding values obtained from correlation mode (cor, symbol).}
\end{figure}

\begin{figure}[H]
\centering
\includegraphics[width=0.9\textwidth]{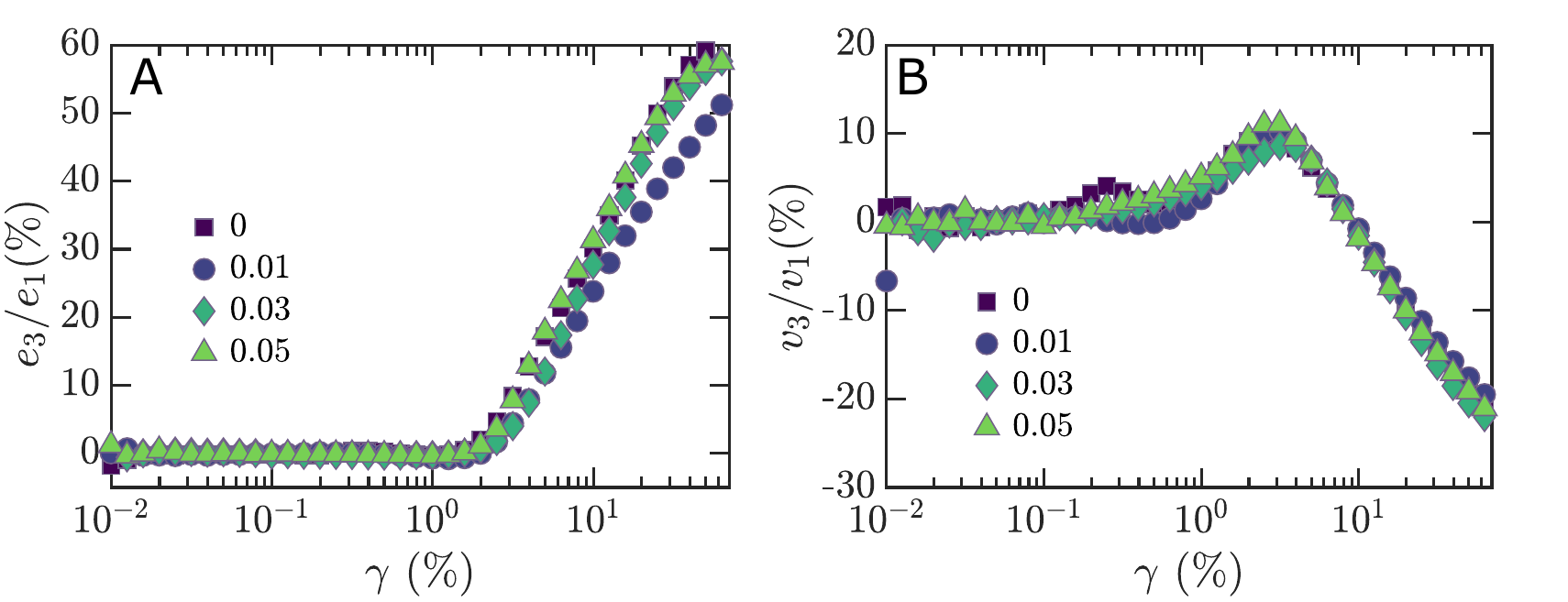}
\caption{\label{fig:e3_v3}Nonlinear rheological behavior from Fourier and Chebyshev analysis. (A) Ratio of elastic third harmonic Chebyshev coefficient to the first harmonic $e_3/e_1$, showcasing the onset and magnitude of strain stiffening behavior. (B) Ratio of viscous third harmonic Chebyshev coefficient to the first harmonic $v_3/v_1$, highlighting shear thickening transitioning to shear thinning with increasing strain amplitude.}
\end{figure}

\begin{figure}[H]
\centering
\includegraphics[width=0.9\textwidth]{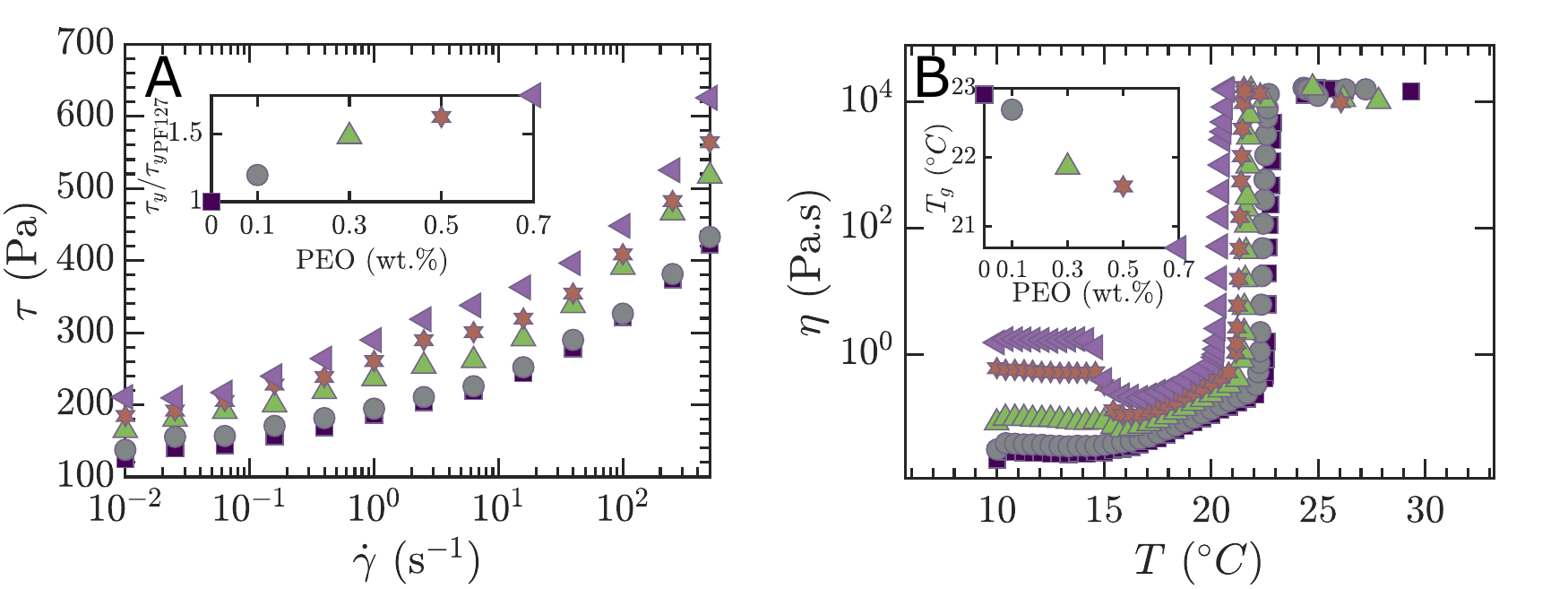}
\caption{\label{fig:PEO_shear}Shear rheology of PF127 with PEO additives. (A) Steady-state flow curves at different PEO concentrations, with the inset highlighting the yield stress $\tau_y$ variation as a function of PEO concentration. (B) Temperature ramp tests conducted at a constant shear stress of 1 Pa and a heating rate of $1^\circ$C/min, showing the effect of PEO concentration on the gelation temperature ($T_g$); the inset presents the resulting shifts in $T_g$.}
\end{figure}

\begin{figure}[H]
\centering
\includegraphics[width=0.5\textwidth]{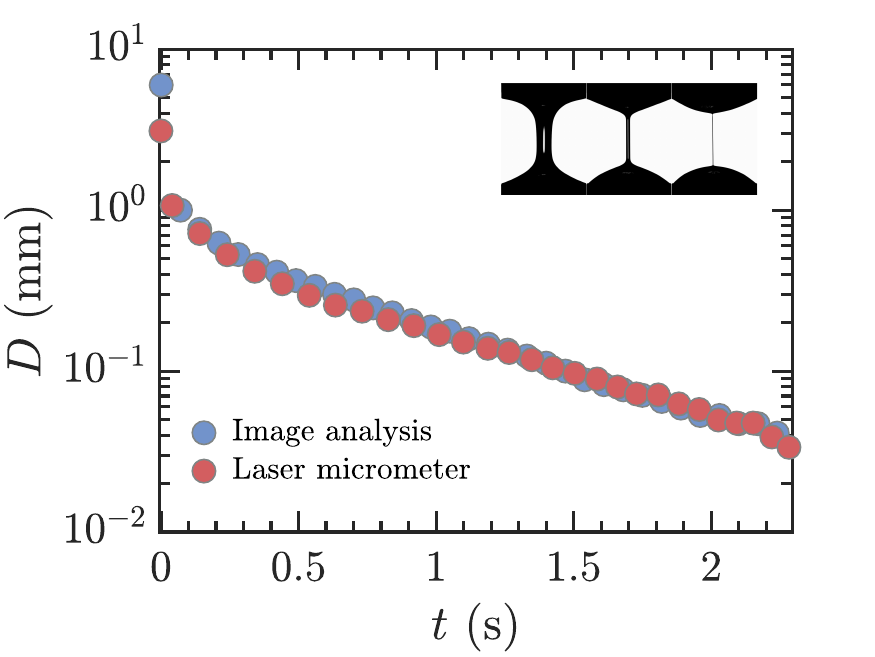}
\caption{\label{fig:CaBER_valid}Validation of our optical measurement system for CaBER analysis. The time evolution of the filament diameter of a $0.5$~wt.$\%$ PEO viscoelastic solution as measured by the built-in laser micrometer of the CaBER against the filament diameter obtained through high-resolution image analysis. The inset presents a sequential time-lapse of the filament thinning captured by our optical system and used in the image analysis.}
\end{figure}

\begin{figure}[H]
\centering
\includegraphics[width=0.85\textwidth]{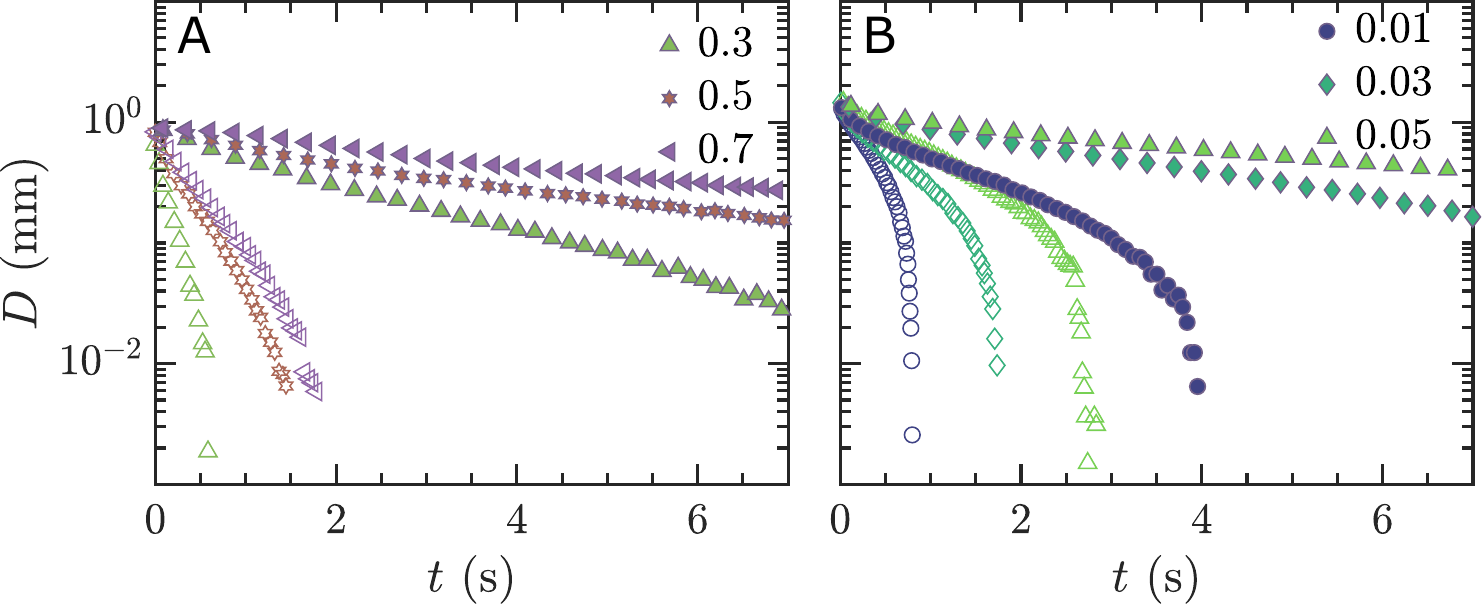}
\caption{\label{fig:CaBER_11C}Time evolution of the filament diameter obtained by the laser micrometer built-in CaBER device at $10^\circ C$ for (A) PEO solutions (empty symbols) and PF127 with PEO additives (filled symbols), and (B) HPAA solutions (empty symbols) and PF127 with HPAA additives (filled symbols). The relaxation times derived from these measurements are summarized in Table ~\ref{table:lambda}.}
\end{figure}

\begin{table}[H]
	\centering
	\caption{\label{table:lambda} Summary of relaxation times. The asterisk ($\ast$) refers to the measurements obtained through image analysis. Note that HPAA additives show similar relaxation times before and after gelation, due to their stable network that minimally interacts with PF127. In contrast, PEO additives exhibit variable relaxation times due to their dynamic interaction with PF127, affecting rheological properties and network stability post-gelation.
	}
	\begin{tabular}{lrrrrrr}
		\\
		Additive (add.) &  (wt.$\%$) & add. solution (10$^\circ$C)& PF127+ add. (10$^\circ$C) & PF127+ add. (26$^\circ$C)$^\ast$ \\ \midrule
		\multirow{3}{*}{PEO} & 0.3  & 56 ms & 797 ms & 10.53 $\pm$ 0.4 ms \\
		& 0.5 & 131 ms & 1.62 s & 48.6 $\pm$ 1.4 ms \\
		& 0.7 & 159 ms & 2.35 s & 65.1 $\pm$ 2.9 ms   \\ \midrule
		\multirow{3}{*}{HPAA} & 0.01 & 110 ms & 443 ms & 316 $\pm$ 21 ms  \\
		& 0.03 & 232 ms & 1.26 s & 1.99 $\pm$ 0.62 s \\
		& 0.05 & 330 ms & 2.19 s & 2.68 $\pm$ 0.36 s \\ 
		\bottomrule
	\end{tabular}
\end{table}

\begin{figure}[H]
\centering
\includegraphics[width=0.75\textwidth]{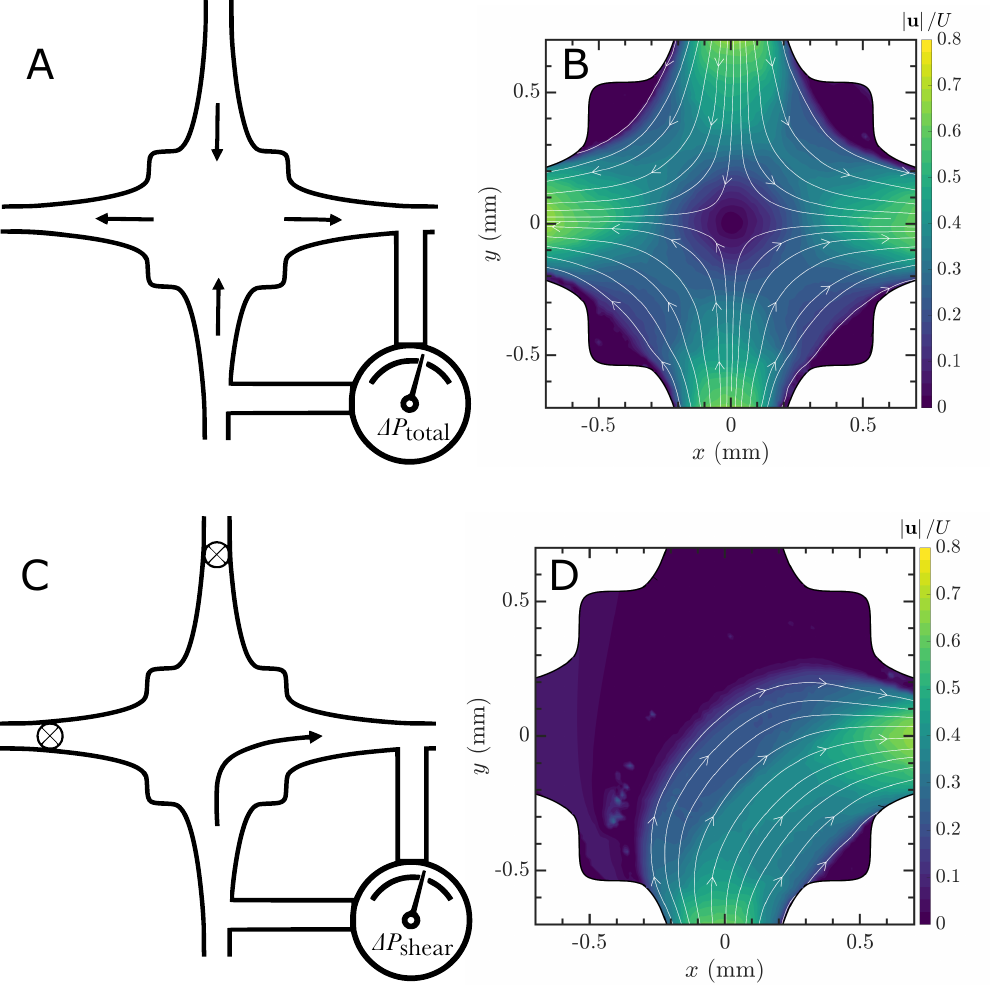}
\caption{\label{fig:P_measure}Pressure difference measurements in the OSCER device. (A) Schematic representation of the total pressure difference $\Delta P_{\text{total}}$ measurement setup with all channels operational under planar elongational flow, maintaining the central stagnation point. (B) Velocity field captured during $\Delta P_{\text{total}}$ measurement. (C) Schematic of shear stress-related pressure difference $\Delta P_{\text{shear}}$ setup, with only one pair of inlet and outlet channels in use, eliminating the stagnation point. (D) Velocity field observed during $\Delta P_{\text{shear}}$ measurement. Velocity fields in (B) and (D) correspond to pure PF127 at a flow rate of $1.6$ mL/min.}
\end{figure}







\end{document}